\documentclass{article}

\PassOptionsToPackage{numbers, compress}{natbib}

\usepackage[preprint]{neurips_2026}

\makeatletter
\renewcommand{\@notice}{}
\renewcommand{\@toptitlebar}{\vskip 0.1in}
\renewcommand{\@bottomtitlebar}{\vskip 0.15in}
\makeatother

\usepackage[utf8]{inputenc}
\usepackage[T1]{fontenc}
\usepackage{hyperref}
\usepackage{url}
\usepackage{booktabs}
\usepackage{amsfonts}
\usepackage{nicefrac}
\usepackage{microtype}
\usepackage[table]{xcolor}
\usepackage{amsmath}
\usepackage{graphicx}


\usepackage{siunitx}

\hypersetup{colorlinks=true, linkcolor={blue!50!black}, citecolor={blue!50!black}, urlcolor={blue!50!black},
  pdftitle={Ablating Design Choices for Protein Ensemble Generators Across Structural Regimes},
  pdfauthor={Eoin Quinn, Oliver Bent}}
\title{Ablating Design Choices for Protein Ensemble Generators Across Structural Regimes}

\author{%
  Eoin Quinn \\
  InstaDeep \\
  London, UK \\
  \texttt{e.quinn@instadeep.com} \\
  \And
  Oliver Bent \\
  InstaDeep \\
  Paris, France \\
  \texttt{o.bent@instadeep.com}  \\
}

\begin{document}

\maketitle
\thispagestyle{plain}

\begin{abstract}
Generative models of protein conformational ensembles inherit much of their design from structure prediction for folded proteins. 
Ordered and disordered proteins, however, occupy distinct structural regimes, and we investigate how design choices hold up across them. 
Using a minimal C$\alpha$ flow-matching model, we ablate choices of architecture, sequence conditioning, rigid-body alignment and training source, and evaluate each variant against experimental SAXS and NMR chemical-shift data partitioned into ordered, disordered, multi-domain and transitional proteins. 
Mixture-of-experts routing, a static pair representation and a disorder probe have no detectable effect, a smaller protein language model performs worse on ordered proteins but not on disordered ones, and alignment in the loss degrades agreement with SAXS. Replacing coarse-grained simulations of disordered proteins with fragment-assembled ensembles improves agreement with chemical shifts at a cost to SAXS, while an equal mixture of the two keeps most of the gain at a smaller cost. This points to the training data, rather than the architecture, as the main constraint.
\end{abstract}

\section{Introduction}
\label{sec:intro}

Structure prediction is largely solved for proteins with a single well-defined fold~\cite{alphafold2,alphafold3}, yet a substantial fraction of biologically active proteins do not adopt one. Intrinsically disordered proteins and regions (IDPs/IDRs), estimated to make up 20--40\% of the human proteome \cite{metapredict}, instead populate heterogeneous conformational ensembles whose flexibility underlies their function \cite{lazar2025,tesei2024}, and many human proteins contain both ordered and disordered regions. Modelling these ensembles underpins applications in drug discovery and protein design \cite{dhar2025,pesce2024}.

While it is tempting to treat the ordered and disordered regimes as endpoints of a continuum, it is more principled to regard them as distinct, as evident in their scaling behaviour ($R_g \sim N^{\nu}$). Ordered proteins adopt a unique compact native fold, and across the proteome they scale with $\nu \approx 1/3$--$2/5$, close to the collapsed-globule limit of a polymer in poor
solvent~\cite{hong2009}. Disordered proteins instead form polymer-like ensembles governed by excluded
volume and conformational entropy, from the
$\theta$-regime ($\nu \approx 1/2$) for typical sequences to the self-avoiding-walk limit
($\nu \approx 0.59$) for highly charged or otherwise expanded ones~\cite{hofmann2012}, determined by sequence-dependent physicochemistry such as net charge, charge patterning and hydropathy~\cite{daspappu2013,tesei2024}.
Multi-domain proteins, in which folded domains are joined by disordered linkers, reinforce this view rather than softening it: each domain or linker sits in one regime and the protein is their composition.

The experimental methods that underpin structure prediction for folded proteins do not transfer to disorder. Crystallography relies on a well-ordered lattice and single-particle cryo-EM reconstruction requires a small number of alignable states, and as a result the PDB is depleted in proteins with disorder. Instead the probes available for disordered proteins measure ensemble-averaged observables, principally small-angle X-ray scattering (SAXS)~\cite{kikhney2015} and NMR chemical shifts (CS)~\cite{camacho2022}. 

As these observables constrain an ensemble only indirectly, ensembles are more often generated by simulation than inferred from data. For molecular dynamics simulations there is an accuracy--efficiency trade-off: all-atom force fields produce high-fidelity trajectories but at an immense computational cost, while coarse-grained methods provide proteome-scale coverage at reduced resolution~\cite{kmiecik2016}. Residue-level models such as CALVADOS \cite{tesei2021,tesei2023} are calibrated top-down against experimental SAXS and PRE data, and so are well tuned for global chain dimensions while sacrificing local geometry. An alternative approach is fragment assembly~\cite{ozenne2012}, which builds ensembles by stitching together short backbone fragments drawn from structural databases.
This method aims for broad coverage and accurate local backbone geometry, but its sampled populations do not follow a Boltzmann distribution.

Recent advances in generative modelling have opened another route, in which models are trained to sample
conformational ensembles directly, amortising over sequence so that no simulation is
required at inference. For disordered regions, idpGAN~\cite{idpgan2023}, idpSAM~\cite{idpsam2024} and STARLING~\cite{starling2025} learn from coarse-grained or implicit-solvent simulations. AlphaFlow and ESMFlow~\cite{alphaflow2024} fine-tune folded-protein structure predictors with flow matching. BioEmu \cite{bioemu} is trained on all-atom molecular dynamics
alongside folded structures. PepTron \cite{invernizzi2025} and IDPFold2 \cite{idpfold2} instead
target the disordered and multi-domain regimes, drawing on proteome-scale ensembles of the
kind described above. Several of these models carry over architectural elements from structure prediction, such as pair representations and structure modules, while SimpleFold~\cite{simplefold2025} shows that folding itself can be learned with general-purpose transformer blocks and no pair representation. 
A complementary line of work predicts ensemble-averaged quantities directly from sequence, without generating an ensemble~\cite{albatross2024,geograph2025}.

In this work we investigate how conformational modelling choices fare across structural regimes, since a choice may matter in one regime and not in another.
To this end we ablate a range of design choices on a minimal C$\alpha$ flow-matching model, and evaluate each variant against experimental SAXS
and chemical shift data partitioned into ordered, disordered, multi-domain and transitional proteins.

\section{Evaluation}
\label{sec:eval}

Since the conformational ensembles available as training data are inherently approximate, we focus our analysis on agreement with experiment. We evaluate against PeptoneBench \cite{invernizzi2025}, which scores generated ensembles against:
\begin{itemize}
    \item SAXS ($n=439$, from the SASBDB~\cite{sasbdb}): constrains global dimensions and shape. It is largely insensitive to fold topology, and so agreement on ordered proteins tests compactness rather than precise fold correctness.
    \item CS ($n=659$, from the BMRB~\cite{bmrb}): constrains local backbone geometry, but carries little information about global chain dimensions.
\end{itemize}
PeptoneBench computes ensemble-averaged observables via forward models, reporting an RMSE normalised by forward-model and experimental uncertainty, so that values near 1 indicate agreement within error. 
The benchmark's default summary statistic is LOWESS-AUC, which integrates the RMSE against a continuous disorder metric, the mean G-score ($0 = \text{ordered}$, $1 = \text{disordered}$).
We score each ensemble as generated, retaining all conformers scored by the forward models and weighting them equally.
We do not apply a physicality filter, so that every model is scored on the same proteins and on its full ensemble (see Appendix~\ref{app:fourcell}).
PeptoneBench can also refit the conformer weights to the measurement, and we report scores after this reweighting in Appendix~\ref{app:reweight}. Reweighting requires the measurement itself, which is unavailable in practice, so ensembles as generated are our primary result.

To resolve performance by regime, we partition PeptoneBench into four cells by G-score. To separate multi-domain proteins we define a block as a contiguous run of $\geq 20$ residues with $G < 0.3$ (rigid) or $G > 0.7$ (flexible), and assign all proteins as follows (cell counts and the scored cohort in Appendix~\ref{app:fourcell}):
\begin{itemize}
    \item multi-domain: contains at least one block of each kind.
    \item ordered / disordered: mean $G < 0.3$ and $G > 0.7$, respectively.
    \item transitional: the remainder, absorbing borderline cases to sharpen the ordered and disordered cells.
\end{itemize}
The ordered and disordered cells isolate the two regimes of Sec.~\ref{sec:intro}, and the multi-domain cell captures their composition within a single chain.

We compare each variant with the base model protein by protein. For each protein we take the ratio of the variant's RMSE to the base model's, each averaged geometrically over training seeds, and summarise a cell by the geometric mean of these ratios, reported as a percentage change. Its 95\% confidence interval accounts for variation across both proteins and training seeds (Appendix~\ref{app:stats}), and we regard a change as negligible when this interval lies within $\pm 5\%$.

In addition, we report the Flory scaling exponent $\nu$ from a population fit of
$R_g \sim N^{\nu}$ within each cell. We compute $R_g$ per conformer from the C$\alpha$ coordinates, weighted by residue mass and averaged over the ensemble, and fit $\nu$ as the least-squares slope of $\log R_g$ against $\log N$ across the proteins of a cell, with a standard error from a bootstrap over proteins. 
Unlike the benchmark scores, $\nu$ does not depend on a
forward model or experimental data, and so provides a consistency check on whether a
model reproduces the chain statistics characteristic of each regime
(Sec.~\ref{sec:intro}).

\section{Base model and ablations}
\label{sec:model}

Our base model (35.7M parameters) is a sequence-conditioned C$\alpha$ flow-matching model predicting a per-residue velocity $v_\theta$ targeting $v_t = x_1 - \epsilon$ under a linear schedule $x_t = t x_1 + (1-t)\epsilon$. Noised coordinates and frozen ESM2-650M embeddings \cite{esm2} are processed by a transformer backbone utilising RoPE \cite{rope} and AdaLN-zero time conditioning \cite{dit}. We train via a per-residue MSE loss with random SO(3) augmentation. 
The C$\alpha$ representation deliberately matches the resolution of the coarse-grained training ensembles, with atomic detail recovered downstream via back-mapping with \texttt{cg2all} \cite{cg2all}. Full architectural, training and sampling details are given in Appendix~\ref{app:arch}. An exploration of mini-batch optimal transport pairing yielded no improvement and is detailed in Appendix~\ref{app:ot}.

We train on data across the structural regimes of Sec.~\ref{sec:intro}. Batches are sampled uniformly from three sources:
high-confidence AFDB structures (pLDDT~$\geq 90$) as single ordered conformers \cite{afdb}, IDRome-CALVADOS simulated ensembles for disordered chains \cite{tesei2024}, and AF-CALVADOS simulations for multi-domain proteins \cite{buelow2026}, in which the whole chain is simulated with CALVADOS and folded domains are restrained towards their AlphaFold2 structures. Additional details are given in Appendix~\ref{app:data}.

Against the minimal base model, we evaluate the following design variants, which are chosen to explore modelling across structural regimes (further details in Appendix~\ref{app:variants}). We train the base model with five random seeds and each variant with three, and sample all models with the same settings.

\paragraph{Sequence conditioning.} While structure prediction for folded proteins leverages the coevolution of spatial contacts, this is less relevant for disordered regions, which rely instead on bulk physicochemical properties. To test how the required sequence-model capacity differs between regimes, we replace ESM2-650M with \mbox{ESM2-8M}.

\paragraph{Pair representation.} Similarly, the dense pair representations central to folded-protein prediction may be redundant for disordered chains. Our base model has no pair bias, and we test adding a static pair representation (Pair) of the kind used by recent ensemble generators such as IDPFold2 \cite{idpfold2}. It combines an outer sum of the ESM2 embeddings with relative sequence positions and binned distances of the noised coordinates, projected to an attention bias in each layer, and is trained with an auxiliary distogram loss on the AFDB structures.

\paragraph{Disorder adaptation.} We consider two variants designed to help the model adapt to ordered versus disordered regions. The first is a frozen per-residue probe on the ESM2 embeddings, trained to predict a pLDDT-derived disorder score and supplied as an additional conditioning signal (Disorder probe). The second is a mixture-of-experts FFN (MoE), motivated by IDPFold2 \cite{idpfold2} (five experts under top-2 routing plus a shared always-on expert; 117M parameters, 69M active per token).

\paragraph{Kabsch alignment.} To handle rotational invariance without explicit equivariance, models often superimpose prediction and target before computing the loss. While this is used to stabilise gradients for rigid domains, for disordered chains extended transient tails can act as lever arms that dominate the superposition, misaligning the rest of the chain and injecting rotational noise into the gradient. Our base model applies no alignment, and we test adding a loss-side Kabsch superposition.

\paragraph{Training source.} CALVADOS is calibrated for global chain dimensions but sacrifices local backbone geometry, whereas fragment assembly inverts this trade-off (Sec.~\ref{sec:intro}). We replace the IDRome-CALVADOS ensembles with IDRome-o~\cite{invernizzi2025}, which is generated by fragment assembly (Fragment), and separately train with half of the disordered data drawn from each source (Fragment~50\%).

\paragraph{Controls.} Finally, we evaluate two controls that deliberately weaken the model, removing RoPE (w/o RoPE) and shrinking it to 11.0M parameters (Size S).

\section{Results}
\label{sec:results}

\begin{figure}[t]
\centering
\includegraphics[width=\textwidth]{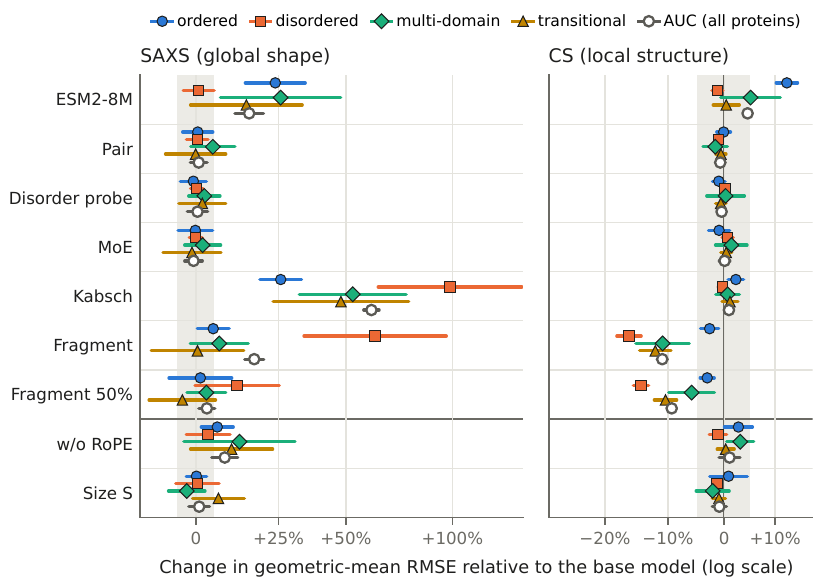}
\caption{Change in geometric-mean RMSE of each variant relative to the base model, per cell and for the benchmark's AUC, with 95\% intervals. Ensembles are scored as generated, and lower is better. The shaded band marks $\pm 5\%$. We regard a change as negligible when its interval lies within the band, and unresolved when its interval contains zero but extends beyond it. The AUC is a single score per run, so its interval reflects variation across training seeds only.}
\label{fig:forest}
\end{figure}

Figure~\ref{fig:forest} shows the change relative to the base model for each variant, per cell and for the benchmark's AUC. Table~\ref{tab:baselines} compares the base model, the Kabsch variant and the two training-source variants with three published ensemble generators, BioEmu \cite{bioemu}, IDPFold2 \cite{idpfold2} and PepTron \cite{invernizzi2025}. Table~\ref{tab:nu} reports the Flory exponent per cell. Excluding frozen sequence encoders, as in our 35.7M count, BioEmu is comparable in size ($\sim$31M) but conditions on AlphaFold2 Evoformer representations computed from a multiple sequence alignment, which is considerably more expensive. IDPFold2 ($\sim$330M total, $\sim$190M active per token) and PepTron ($\sim$690M, a fine-tuned ESMFold) are roughly an order of magnitude larger, so comparisons with all three are indicative rather than controlled.

\begin{table}[t]
\centering
\footnotesize
\setlength{\tabcolsep}{1.6pt}
\caption{Absolute scores for the three baselines, the base model, the Kabsch variant and the two training-source variants. Each entry is the geometric-mean normalised RMSE over the proteins in a cell, with ensembles scored as generated and without the default physicality filter (Appendix~\ref{app:fourcell}). The base model and variants are first averaged geometrically over training seeds. AUC is the benchmark's LOWESS-AUC. Lower is better. Cells contain 313/34/42/42 (SAXS) and 237/220/40/150 (CS) proteins.}
\label{tab:baselines}
\begin{tabular}{lccccc@{\hskip 8pt}ccccc}
\toprule
& \multicolumn{5}{c}{SAXS} & \multicolumn{5}{c}{CS} \\
\cmidrule(lr){2-6}\cmidrule(lr){7-11}
& ordered & disordered & multi-domain & transitional & AUC & ordered & disordered & multi-domain & transitional & AUC \\
\midrule
BioEmu & 2.06 & 3.50 & 3.37 & 2.54 & 2.57 & 1.04 & 1.48 & 1.25 & 1.22 & 1.24 \\
IDPFold2 & 2.51 & 2.43 & 3.92 & 3.21 & 2.68 & 1.10 & 1.24 & 1.14 & 1.21 & 1.17 \\
PepTron & 2.23 & 4.60 & 3.66 & 2.60 & 3.01 & 1.03 & 1.80 & 1.41 & 1.46 & 1.42 \\
\midrule
Base & 2.42 & 2.01 & 3.06 & 2.90 & 2.29 & 1.24 & 1.40 & 1.34 & 1.31 & 1.31 \\
\midrule
Kabsch & 3.04 & 3.99 & 4.67 & 4.29 & 3.68 & 1.27 & 1.39 & 1.35 & 1.33 & 1.33 \\
Fragment & 2.53 & 3.26 & 3.25 & 2.91 & 2.68 & 1.21 & 1.17 & 1.20 & 1.15 & 1.17 \\
Fragment 50\% & 2.45 & 2.24 & 3.14 & 2.80 & 2.36 & 1.20 & 1.20 & 1.26 & 1.18 & 1.19 \\
\bottomrule
\end{tabular}
\end{table}

\begin{table}[t]
\centering
\footnotesize
\setlength{\tabcolsep}{5pt}
\caption{Flory scaling exponent $\nu$ from a population fit of $R_g \sim N^{\nu}$, with bootstrap SE, for the models of Table~\ref{tab:baselines}. The multi-domain cell is omitted, as its narrow length range makes $\nu$ unreliable. The other variants are within error of the base model in these cells.}
\label{tab:nu}
\begin{tabular}{lccc}
\toprule
 & ordered & disordered & transitional \\
\midrule
BioEmu & 0.41\,$\pm$\,0.02 & 0.69\,$\pm$\,0.02 & 0.48\,$\pm$\,0.02 \\
IDPFold2 & 0.34\,$\pm$\,0.01 & 0.56\,$\pm$\,0.02 & 0.42\,$\pm$\,0.02 \\
PepTron & 0.38\,$\pm$\,0.01 & 0.50\,$\pm$\,0.02 & 0.45\,$\pm$\,0.02 \\
\midrule
Base & 0.34\,$\pm$\,0.01 & 0.54\,$\pm$\,0.01 & 0.45\,$\pm$\,0.02 \\
\midrule
Kabsch & 0.35\,$\pm$\,0.02 & 0.58\,$\pm$\,0.01 & 0.48\,$\pm$\,0.02 \\
Fragment & 0.33\,$\pm$\,0.02 & 0.61\,$\pm$\,0.01 & 0.49\,$\pm$\,0.02 \\
Fragment 50\% & 0.34\,$\pm$\,0.01 & 0.55\,$\pm$\,0.01 & 0.47\,$\pm$\,0.02 \\
\bottomrule
\end{tabular}
\end{table}

The base model is a reasonable reference (Table~\ref{tab:baselines}). 
It is competitive with the three baselines on SAXS, mainly through disordered proteins, but trails IDPFold2 on chemical shifts in every cell, and all three baselines on ordered proteins. 
Its Flory exponents follow the expected ordering, from 0.34 for ordered to 0.54 for disordered proteins (Table~\ref{tab:nu}).

Reducing the size of the protein language model degrades performance on ordered and multi-domain proteins but not on disordered ones (Figure~\ref{fig:forest}). With \mbox{ESM2-8M} in place of ESM2-650M, ordered proteins are worse on both observables (SAXS $+24\%$, chemical shifts $+12\%$), as are multi-domain proteins on SAXS ($+26\%$), while disordered proteins show no detectable change, in line with the coevolutionary argument of Sec.~\ref{sec:model}.

None of the added architectural components has a detectable effect. The disorder probe, the mixture of experts, which more than triples the parameter count, and the pair representation, which adds a quadratic pair track, stay within $\pm 5\%$ of the base model wherever the data can resolve it. Only the small multi-domain and transitional SAXS cells are too noisy to rule out a 5\% change.

Loss-side Kabsch alignment degrades SAXS in every cell, from $+26\%$ on ordered proteins to $+99\%$ on disordered ones, and leaves chemical shifts unaffected. It is worse even on ordered proteins, where the gradient stabilisation that motivates it should help most, and the largest effect, on disordered proteins, is consistent with the lever-arm mechanism of Sec.~\ref{sec:model}.
The degradation persists after reweighting (Appendix~\ref{app:reweight}), suggesting that the ensembles contain too few conformers compatible with the measurement.
In the disordered cell its Flory exponent rises above the base model's ($\nu = 0.58$ against $0.54$), close to the self-avoiding-walk value, indicating over-expansion.

Only the training source improves either observable. Replacing the CALVADOS ensembles with fragment-assembled ones improves chemical shifts by 11--16\% outside the ordered cell, bringing the chemical-shift AUC level with IDPFold2's, but worsens SAXS on disordered proteins by 62\% and raises their Flory exponent to 0.61. Fragment~50\% keeps most of the chemical-shift gain ($-9\%$ against $-11\%$ in AUC) at a fraction of the SAXS cost, with the SAXS AUC only slightly worse.

Even deliberately weakening the model changes little. The small model is a third the size of the base model and shows no detectable effect, while removing RoPE costs only a few percent, mostly on SAXS (AUC~$+8\%$). 
\section{Conclusion}
\label{sec:conclusion}

The ablations suggest that, at this scale, the model is limited by its training data rather than its architecture. Adding architectural components has no detectable effect, and even deliberately weakening the model costs little. The choices that do matter act on particular regimes or observables, which only an evaluation resolved by regime can show. Of the two premised on a stable native fold, the size of the protein language model matters only for folded proteins, an effect the chemical-shift AUC barely registers, while alignment in the loss degrades agreement with SAXS in every regime. The source of the disordered training data trades local against global accuracy, a trade-off that an equal mixture of the two sources softens. Only the training source improves either observable, so further progress will need training sources that are accurate at both local and global scales and across structural regimes.

\paragraph{Limitations.}
Our training data come from coarse-grained simulation, fragment assembly and predicted structures, all of them approximations, and CALVADOS was calibrated against experimental data that may overlap with PeptoneDB-SAXS~\cite{buelow2026}. 
Our C$\alpha$ model is compared with experiment only after back-mapping, whose error we do not quantify. 
The ordered data are also restricted to confidently predicted structures (pLDDT~$\geq 90$), which leaves partially ordered regions underrepresented.
Our conclusions about architecture are drawn at a single scale, with ordered, disordered and multi-domain data in fixed equal proportion. Finally, the evaluation rests on two observables. In particular, SAXS on ordered proteins tests compactness rather than fold correctness, and with three seeds per variant, the small non-ordered SAXS cells (34--42 proteins) leave several comparisons unresolved.

\bibliographystyle{unsrtnat}
\bibliography{references}

\clearpage
\appendix

\section{Architecture and training}
\label{app:arch}

Table~\ref{tab:hparams} lists the base model's architecture and training hyperparameters.

\begin{table}[h]
\centering
\footnotesize
\caption{Base model architecture and training hyperparameters.}
\label{tab:hparams}
\begin{tabular}{ll}
\toprule
\multicolumn{2}{l}{\emph{Architecture} (35.7M trainable; frozen ESM2-650M excluded)} \\
\midrule
hidden size / layers / heads / head dim & 384 / 12 / 6 / 64 \\
feed-forward & SwiGLU, $\tfrac{8}{3}$ expansion ($\approx$1024) \\
per-block norm / gating & RMSNorm on Q,\,K; sigmoid attention-output gate~\cite{qiu2025gated} \\
conditioning & AdaLN-zero (time) \\
positional encoding & RoPE, half of head dims rotated \\
coordinate scaler & divide by 1.5\,nm, centred \\
\midrule
\multicolumn{2}{l}{\emph{Training}} \\
\midrule
optimiser & AdamW ($\beta=0.9,\,0.999$; weight decay 0.01) \\
learning rate & $5\times10^{-4}$; 5k-step linear warmup, cosine decay to 1\% \\
batch size / steps / precision & 128 / 100k / bf16-mixed \\
grad clip (norm) / EMA decay & 2.0 / 0.9999 \\
timestep sampling & logit-normal ($\mu = 0.5$, $s = 1$; median $t\approx0.62$), 10\% uniform \\
\bottomrule
\end{tabular}
\end{table}

\paragraph{Sampling and back-mapping.} Ensembles are generated by integrating the learned velocity from $t{=}0$ to $t{=}1$ with a stochastic Euler solver (diffusion coefficient $\lambda(t)=0.5\,\sigma(t)^2$, with $\sigma(t) = 1-t$ the noise scale of the linear schedule) over 100 steps, re-centring at each step, and draw 100 conformers per protein. C$\alpha$ ensembles are mapped to all-atom structures with \texttt{cg2all}~\cite{cg2all}.

\section{Data}
\label{app:data}

\paragraph{Sources and preprocessing.} We train on three datasets at C$\alpha$ resolution (one bead per residue), restricting to sequences of $30$--$510$ residues:
\begin{itemize}
    \item \textbf{AFDB} \cite{afdb} (ordered): Foldcomp \texttt{afdb\_rep\_v4}, mean pLDDT~$\geq 90$, a length-stratified subsample of $100{,}000$ proteins with one conformer each.
    \item \textbf{IDRome-CALVADOS} \cite{tesei2024,tesei2021,tesei2023} (disordered): Zenodo \texttt{10251736}.
    \item \textbf{AF-CALVADOS} \cite{buelow2026} (multi-domain): cytosolic subset, from ERDA \url{https://sid.erda.dk/sharelink/FxaWhNrQgp}.
\end{itemize}

\paragraph{Batch composition.} Minibatches are stratified across the three datasets in equal $1{:}1{:}1$ proportion. 

\paragraph{Data splits.} We cluster the pooled sequences of all training datasets with MMseqs2 at 40\% identity and assign each cluster whole to one split ($\approx$90/5/5 train/validation/test). Clusters containing a PeptoneBench sequence are forced into the test split.

\paragraph{AF-CALVADOS C$\alpha$ restoration.} The source trajectories represent ordered domains by per-residue centre-of-mass beads, which lack physical C$\alpha$ geometry. We restore it in each frame. Within every rigid domain (contiguous high-pLDDT residues, confirmed by PAE) we Kabsch-align the AlphaFold2 structure onto the domain's COM beads and take its C$\alpha$ coordinates, leaving disordered residues as their CALVADOS beads. A short healing pass then repairs the C$\alpha$--C$\alpha$ bonds at domain--linker junctions.

\section{Variant details}
\label{app:variants}

This appendix gives details of the variants not fully specified in Sec.~\ref{sec:model}.

\subsection{Disorder probe}
\label{app:disorder_probe}
The probe is a two-layer MLP ($D_\text{esm}\!\to\!D_\text{esm}\!\to\!1$, SiLU nonlinearity, 0.1 dropout) that maps each ESM2 per-residue embedding to a scalar disorder logit. It is trained separately on AFDB structures to predict $D_i = 1 - \text{pLDDT}_i/100$ under a masked binary cross-entropy loss, with proteins sampled in equal proportion across four pLDDT bands ($[0,50)$, $[50,70)$, $[70,90)$, $[90,100)$). During generator training the probe is frozen. Its per-residue score is Fourier-embedded (32 dimensions, zero-initialised so the pathway starts inactive) and concatenated with the ESM2 embeddings before the sequence projection.

\subsection{Mixture of experts}
The MoE variant replaces the feed-forward block with five routed experts under top-2 routing plus one always-on shared expert. Routed and shared outputs are combined as $(\text{shared} + 2\,\text{routed})/3$, and a Switch-style load-balancing auxiliary loss (coefficient $0.01$) discourages routing collapse. This raises the trainable parameter count to 117M.

\subsection{Static pair representation}
The pair track has channel width 128 and is initialised by combining an outer sum of the ESM2 embeddings, a relative-position encoding ($|i-j|$ clamped at $\pm32$, 65 bins), and a binned C$\alpha$-distance seed of the noised coordinates (32 bins over $0$--$4$\,nm, $0.125$\,nm wide). The initialisation is time-conditioned through an AdaLN layer, zero-initialised so that the modulation starts as the identity. It is projected to a per-head attention bias at every transformer layer. A distogram head on the pair track is trained with an auxiliary cross-entropy loss (weight 0.1, 64 bins over 0.2--2.2\,nm) on AFDB examples only. The static variant applies no triangle-multiplicative updates (zero pair-update blocks).

\subsection{Kabsch alignment}
The Kabsch variant rotates the ground truth $x_1$ onto the model's current estimate of the clean structure, obtained from the predicted velocity, and re-derives the velocity target as $(R x_1 - x_t)/(1-t)$, so that the implied clean structure is the rigidly rotated $R x_1$. The rotation is computed without gradients.

\subsection{Training source}
Fragment replaces IDRome-CALVADOS with IDRome-o \cite{invernizzi2025} in the disordered third of each batch. Fragment~50\% fills that third equally from the two sources. The ordered (AFDB) and multi-domain (\mbox{AF-CALVADOS}) sources are unchanged.

\subsection{Size S}
The Size S control reduces the hidden size to 256, depth to 8 layers, and heads to 4 (from the base $384/12/6$), giving 11.0M trainable parameters.

\section{On mini-batch optimal-transport pairing}
\label{app:ot}

Mini-batch optimal-transport (OT) pairing matches each data sample $x_1$ with a noise sample $\epsilon$ within a batch to shorten the average flow path and reduce the variance of the velocity field~\cite{tong2024}. We tested two cheap approximations to exact OT, and neither helped.

\paragraph{Pair-by-scale.} For each data sample we compute its RMS scale, normalised by the RMSD of its parent ensemble to account for how flexible the ensemble is. We then sort the data and noise samples by scale and pair them by rank. Isotropic Gaussian noise in $3L$ dimensions, however, concentrates on a thin shell where all samples have nearly the same scale, so the sort barely changes the pairing.

\paragraph{Pair-by-rotation.} For each pair $(x_1, \epsilon)$ we use SO(3) invariance to rotate $x_1$ into the frame of $\epsilon$ by Kabsch alignment. This made training worse, for which we see three likely reasons. The gain is small to begin with. Two random configurations in $3L$ dimensions are nearly orthogonal, and a rotation can only exploit their $3 \times 3$ cross-covariance, whose singular values grow as $\sqrt{L}$. The alignment therefore shortens the squared transport distance by $O(\sqrt{L})$ out of $O(L)$, a fractional gain of order $1/\sqrt{L}$, or roughly 10\% for a 100-residue chain. Training on aligned pairs also confines the velocity field to a thin, rotationally aligned manifold, from which the trajectory drifts at inference and errors compound. Translational drift is removed by re-centring at each step, but rotational drift is not. Finally, as with loss-side Kabsch alignment, extended tails dominate the superposition and bias it towards matching the tails.

\section{Four-cell partition of PeptoneBench}
\label{app:fourcell}

PeptoneBench \cite{invernizzi2025} benchmarks protein conformational-ensemble generators against experimental observables. We use its two primary datasets: PeptoneDB-CS ($n=659$ proteins with NMR chemical shifts from BMRB) and PeptoneDB-SAXS ($n=439$ proteins with SAXS profiles from SASBDB). Ensemble-averaged observables are computed via forward models and compared to experiment as a normalised RMSE accounting for forward-model and experimental uncertainty. Per-protein disorder is quantified by a mean G-score in $[0,1]$ ($0 = \text{ordered}$, $1 = \text{disordered}$), derived from secondary chemical shifts for CS proteins and predicted from sequence for SAXS proteins. The summary score for each observable is the LOWESS-AUC, obtained by plotting per-protein RMSE against mean G-score, smoothing with LOWESS and integrating over the range covered by the data.

We partition each dataset into four cells by G-score, as follows. To identify multi-domain proteins, we define blocks as contiguous runs of $\geq 20$ residues with $G < 0.3$ (rigid blocks) or $G > 0.7$ (flexible blocks). The cells are mutually exclusive. Multi-domain detection takes priority, and the remaining proteins are assigned by mean G-score.
\begin{itemize}
    \item \textbf{multi-domain}: $\geq 1$ rigid block and $\geq 1$ flexible block (regardless of mean $G$).
    \item \textbf{ordered}: not multi-domain, $G < 0.3$.
    \item \textbf{disordered}: not multi-domain, $G > 0.7$.
    \item \textbf{transitional}: not multi-domain, $G \in [0.3, 0.7]$.
\end{itemize}
The \textit{transitional} cell is introduced to sharpen the ordered and disordered cells. It may contain borderline cases, such as molten-globule-like proteins (uniform partial flexibility) or mosaic chains (order-disorder transitions at sub-block scales).

Cell counts per dataset, computed under the default thresholds $(0.3, 0.7)$ and 20-residue block-length floor, are below.

\begin{center}
\begin{tabular}{lcccc}
\toprule
observable & ordered & disordered & multi-domain & transitional \\
\midrule
CS ($n=659$)   & 243 & 220 & 42 & 154 \\
SAXS ($n=439$) & 319 & 34  & 43 & 43  \\
\bottomrule
\end{tabular}
\end{center}

\paragraph{Scoring cohort.} All results use the proteins scored by every model, including the baselines, which gives 431 SAXS and 647 CS proteins (ordered/disordered/multi-domain/transitional: 313/34/42/42 and 237/220/40/150). The shortfall from the full partition is due to proteins absent from the published BioEmu and IDPFold2 ensembles. We additionally exclude one SAXS protein (\texttt{SASDRS5}), whose PeptoneBench-distributed scattering curve is sign-inverted and cannot be fit.

\paragraph{Pipeline calibration.} With the benchmark's default settings, our re-implementation reproduces PeptoneBench's published baseline summary scores. After reweighting, a small residual remains, arising from the unseeded random initialisation of the benchmark's reweighting optimiser.

\paragraph{Physicality filter.} PeptoneBench by default discards conformers that fail the physicality checks of BioEmu~\cite{bioemu}. We apply no filter to any model, so that every model is scored on the same proteins and on its full ensemble. Applying the filter reverses none of the variant comparisons in Sec.~\ref{sec:results}. It removes a larger share of conformers from our C$\alpha$ ensembles than from the baselines', so the absolute comparison with the baselines (Table~\ref{tab:baselines}) should be read as indicative.

\section{Statistical comparison}
\label{app:stats}

We compare each variant $V$ with the base model $B$ by combining a paired comparison over proteins with a Welch comparison over training seeds. Let $r_{msp}$ denote the normalised RMSE of model $m$, trained with seed $s$, on protein $p$. The variant has $k_V = 3$ seeds and the base model $k_B = 5$.

\paragraph{Effect.} For each protein in a cell we take the difference of the seed-averaged log RMSEs,
\[
d_p = \frac{1}{k_V}\sum_{s} \log r_{Vsp} \;-\; \frac{1}{k_B}\sum_{s} \log r_{Bsp},
\]
and average it over the $n$ proteins to give $\bar d$. Exponentiating $\bar d$ gives the geometric mean of the per-protein ratios of Sec.~\ref{sec:eval}, which we report as a percentage change, $100\,(e^{\bar d}-1)\%$.

\paragraph{Uncertainty.} The standard error of $\bar d$ is
\[
\mathrm{SE}^2 = \frac{S_d^2}{n} + \frac{S_V^2}{k_V} + \frac{S_B^2}{k_B},
\]
where $S_d^2$ is the variance of $d_p$ across proteins, and $S_V^2$ and $S_B^2$ are the variances across seeds of each model's log RMSE averaged over the cell. The 95\% interval is $\bar d \pm t_{\mathrm{df}}\,\mathrm{SE}$, with $t_{\mathrm{df}}$ from a $t$ distribution with
\[
\mathrm{df} = \mathrm{SE}^4 \bigg/ \left[ \frac{(S_d^2/n)^2}{n-1} + \frac{(S_V^2/k_V)^2}{k_V-1} + \frac{(S_B^2/k_B)^2}{k_B-1} \right]
\]
degrees of freedom, as in Welch's test.

\paragraph{Reading.} Intervals that contain zero but extend beyond $\pm 5\%$ are unresolved. As a check, we split the base model's seeds into two groups (three and two) and compared them as if they were different models, and no comparison read as better or worse.

\section{Scores after reweighting}
\label{app:reweight}

PeptoneBench also scores each ensemble after reweighting it towards the measurement. The conformer weights are refitted by maximum entropy, keeping an effective sample size of at least 10 conformers out of about 100. Because the reweighted score uses the measurement it is evaluated against, it indicates whether an ensemble contains conformers compatible with the measurement, not whether it generates them in the right proportions.

Reweighting lowers every score and narrows the differences between models on SAXS (Table~\ref{tab:baselines_rw}). On SAXS the base model is then competitive with the three baselines in AUC and in every cell except the transitional one.

\begin{table}[h]
\centering
\footnotesize
\setlength{\tabcolsep}{1.6pt}
\caption{As Table~\ref{tab:baselines}, after PeptoneBench's reweighting. Lower is better.}
\label{tab:baselines_rw}
\begin{tabular}{lccccc@{\hskip 8pt}ccccc}
\toprule
& \multicolumn{5}{c}{SAXS} & \multicolumn{5}{c}{CS} \\
\cmidrule(lr){2-6}\cmidrule(lr){7-11}
& ordered & disordered & multi-domain & transitional & AUC & ordered & disordered & multi-domain & transitional & AUC \\
\midrule
BioEmu & 1.39 & 1.36 & 1.48 & 1.27 & 1.21 & 0.93 & 1.10 & 1.05 & 0.90 & 0.99 \\
IDPFold2 & 1.39 & 1.15 & 1.45 & 1.34 & 1.15 & 0.96 & 0.89 & 0.91 & 0.84 & 0.91 \\
PepTron & 1.57 & 1.76 & 1.84 & 1.48 & 1.47 & 0.93 & 1.33 & 1.08 & 1.08 & 1.11 \\
\midrule
Base & 1.36 & 1.15 & 1.32 & 1.41 & 1.14 & 1.12 & 1.18 & 1.16 & 1.04 & 1.14 \\
\midrule
Kabsch & 1.75 & 1.70 & 2.27 & 2.03 & 1.69 & 1.15 & 1.18 & 1.18 & 1.07 & 1.15 \\
Fragment & 1.37 & 1.36 & 1.49 & 1.47 & 1.20 & 1.06 & 0.90 & 1.01 & 0.87 & 0.97 \\
Fragment 50\% & 1.37 & 1.12 & 1.36 & 1.40 & 1.13 & 1.06 & 0.94 & 1.05 & 0.89 & 0.99 \\
\bottomrule
\end{tabular}
\end{table}

\begin{figure}[h]
\centering
\includegraphics[width=\textwidth]{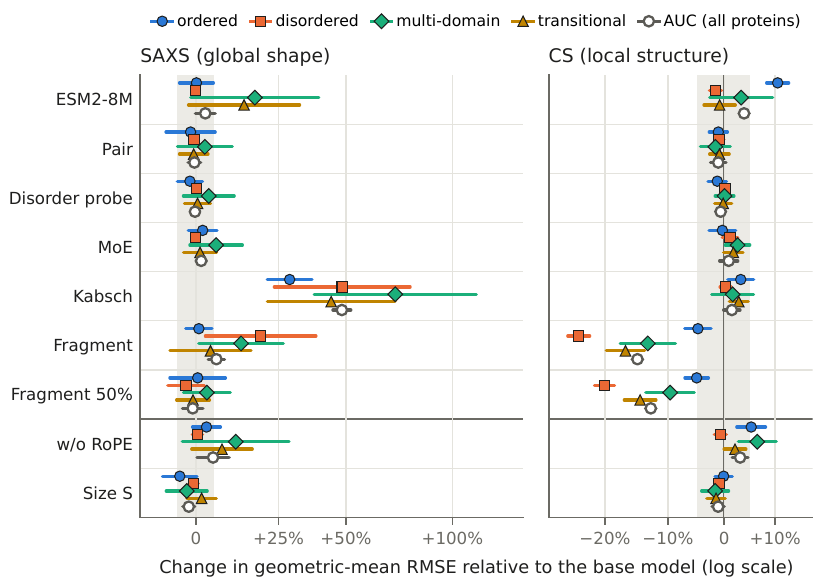}
\caption{As Figure~\ref{fig:forest}, after PeptoneBench's reweighting. Points give the change in geometric-mean RMSE relative to the base model, per cell and for the benchmark's AUC, with 95\% intervals. Lower is better, and the shaded band marks $\pm 5\%$. The AUC interval reflects variation across training seeds only.}
\label{fig:effects_rw}
\end{figure}

Most comparisons with the base model read as they do without reweighting (Figure~\ref{fig:effects_rw}). The SAXS degradation of \mbox{ESM2-8M} on ordered proteins disappears ($+24\%$ to $0\%$), consistent with suitable conformers being present but under-weighted, while its chemical-shift deficit on ordered proteins remains ($+11\%$). Kabsch remains worse in every SAXS cell ($+29\%$ to $+71\%$), although less so on disordered proteins ($+99\%$ to $+48\%$), suggesting that its ensembles contain too few conformers compatible with the measurement for reweighting to recover. For Fragment, the SAXS degradation on disordered proteins shrinks ($+62\%$ to $+19\%$) and the chemical-shift improvement grows ($-16\%$ to $-24\%$). For Fragment~50\%, the SAXS change on disordered proteins moves from $+12\%$ to $-3\%$ (both unresolved), and chemical shifts improve in every cell, including ordered proteins ($-5\%$).

\end{document}